\documentclass[12pt,a4paper,english]{article}
\usepackage[authoryear]{natbib}
\usepackage{graphicx}
\usepackage{isorot}
\usepackage{geometry}
\newcommand{\degr}{$^{\circ}$}  
\newcommand{\ignore}[1]{}
\newcommand{\ex}{$\times$} 

\makeatother
\begin{document}

\title{{A review of wildland fire spread modelling, 1990-present}\\
      { 1: Physical and quasi-physical models}}

\author{A.L. Sullivan}

\maketitle
{\small

\begin{center}

Ensis\footnote{A CSIRO/Scion Joint Venture} Bushfire Research \footnote{current address: Department of Theoretical Physics,\\
Research School of Physical Sciences and Engineering,\\
The Australian National University, Canberra 0200, Australia. }\\
PO Box E4008, Kingston, ACT 2604, Australia

email: Andrew.Sullivan@ensisjv.com or als105@rsphysse.anu.edu.au

phone: +61 2 6125 1693, fax: +61 2 6125 4676

\end{center}}

\begin{center}version 3.0 \end{center}

\setcounter{secnumdepth}{0}
\setlength\parskip{\bigskipamount}
\setlength\parindent{0pt}

\begin{abstract}
In recent years, advances in computational power and spatial data analysis
(GIS, remote sensing, etc) have led to an increase in attempts to model the
spread and behaviour of wildland fires across the landscape. This series of
review papers endeavours to critically and comprehensively review all types of
surface fire spread models developed since 1990. This paper reviews models of a
physical or quasi-physical nature. These models are based on the fundamental
chemistry and/or physics of combustion and fire spread. Other papers in the
series review models of an empirical or quasi-empirical nature, and
mathematical analogues and simulation models. Many models are extensions or
refinements of models developed before 1990. Where this is the case, these
models are also discussed but much less comprehensively.
\end{abstract}

\section{Introduction}

\subsection{History}

The field of wildland fire behaviour modelling has been active since the 1920s.
The work of \citet{Hawley1926} and \citet{Gisborne1927,Gisborne1929} pioneered
the notion that understanding of the phenomenon of wildland fire and the
prediction of the danger posed by a fire could be gained through measurement
and observation and theoretical considerations of the factors that might
influence such fires. Despite the fact that the field has suffered from a lack
of readily achievable goals and consistent funding \citep{Williams1982}, the
pioneering work by those most closely affected by wildland fire--the foresters
and other land managers--has led to a broad framework of understanding of
wildland fire behaviour that has enabled the construction of operational models
of fire behaviour and spread that, while not perfect for every situation, at
least get the job done.

In the late 1930s and early 1940s, \citet{Curry1938,Curry1940}, and
\citet{Fons1946} brought a rigorous physical approach to the measurement
and modelling of the behaviour of wildland fires. In the early 1950s,
formal research initiatives by Federal and State Government forestry
agencies commenced concerted efforts to build fire danger rating systems
that embodied a fire behaviour prediction component in order to better
prepare for fire events. In the US this was through the Federal US
Forest Service and through State agencies; in Canada this was the
Canadian Forest Service; in Australia this was through the Commonwealth
Forestry and Timber Bureau in conjunction with various state authorities.

In the 1950s and 60s, spurred on by incentives from defense budgets,
considerable effort was expended exploring the effects of mass bombing (such as
occurred in Dresden or Hamburg, Germany, during World War Two) and the
collateral incendiary effects of nuclear weapons
\citep{Lawson1954b,Rogers1963}. This research effort was closely related to
large forest or conflagration fires and had the spin-off of bringing additional
research capacity into the field \citep{Chandler1963}. This resulted in an
unprecedented boom in the research of wildland fires. The late 1960s saw a
veritable explosion of research publications connected to wildland fire that
dominated the fields of combustion and flame research for some years.

The 1970s saw a dwindling of research interest from defense organisations and
by the 1980s, research into the behaviour of wildland fires returned to those
that had direct interest in the understanding and control of such phenomena. By
the 1980s, it was of occasional interest to journeyman mathematicians and
physicists on their way to bigger, and more achievable, goals.

An increase in the capabilities of remote sensing, geographical information
systems and computing power during the 1990s resulted in a revival in the
interest of fire behaviour modelling, this time applied to the prediction of
fire spread across the landscape.

\subsection{Background}

This series of review papers endeavours to comprehensively and critically
review the extensive range of modelling work that has been conducted in recent
years. The range of methods that have been undertaken over the years represents
a continuous spectrum of possible modelling \citep{Karplus1977a}, ranging from
the purely physical (those that are based on fundamental understanding of the
physics and chemistry involved in the combustion of biomass fuel and behaviour
of a wildland fire) through to the purely empirical (those that have been based
on phenomenological description or statistical regression of observed fire
behaviour). In between is a continuous meld of approaches from one end of the
spectrum or the other. \citet{Weber1991a} in his comprehensive review of
physical wildland fire modelling proposed a system by which models were
described as physical, empirical or statistical, depending on whether they
account for different modes of heat transfer, make no distinction between
different heat transfer modes, or involve no physics at all. \citet{Pastor2003}
proposed descriptions of theoretical, empirical and semi-empirical, again
depending on whether the model was based on purely physical understanding, of a
statistical nature with no physical understanding, or a combination of both.
\citet{Grishin1997} divided models into two classes, deterministic or
stochastic-statistical. However, these schemes are rather limited given the
combination of possible approaches and, given that describing a model as
semi-empirical or semi-physical is a `glass half-full or half-empty' subjective
issue, a more comprehensive and complete convention was required.

Thus, this review series is divided into three broad categories: physical and
quasi-physical models; empirical and quasi-empirical models; and simulation and
mathematical analogue models. In this context, a physical model is one that
attempts to represent both the physics and chemistry of fire spread; a
quasi-physical model attempts to represent only the physics. An empirical model
is one that contains no physical basis at all (generally only statistical in
nature), a quasi-empirical model is one that uses some form of physical
framework upon which to base the statistical modelling chosen. Empirical and
quasi-empirical models are further subdivided into field-based and
laboratory-based. Simulation models are those that implement the preceding
types of models in a simulation rather than modelling context. Mathematical
analogue models are those that utilise a mathematical precept rather than a
physical one for the modelling of the spread of wildland fire.

Since 1990, there has been rapid development in the field of spatial data
analysis, e.g. geographic information systems and remote sensing. Following
this, and the fact that there has not been a comprehensive critical review of
fire behaviour modelling since \citet{Weber1991a}, I have limited this review
to works published since 1990. However, as much of the work that will be
discussed derives or continues from work carried out prior to 1990, such work
will be included much less comprehensively in order to provide context.

\subsection{Previous reviews}

Many of the reviews that have been published in recent years have been for
audiences other than wildland fire researchers and conducted by people without
an established background in the field. Indeed, many of the reviews read like
purchase notes by people shopping around for the best fire spread model to
implement in their part of the world for their particular purpose. Recent
reviews (e.g. \citet{Perry1998,Pastor2003}; etc), while endeavouring to be
comprehensive, have offered only superficial and cursory inspections of the
models presented. \citet{Morvan2004a} takes a different line by analysing a
much broader spectrum of models in some detail and concludes that no single
approach is going to be suitable for all uses.

While the recent reviews provide an overview of the models and approaches that
have been undertaken around the world, mention must be made of significant
reviews published much earlier that discussed the processes in wildland fire
propagation themselves. Foremost is the work of \citet{Williams1982}, which
comprehensively covers the phenomenology of both wildland and urban fire, the
physics and chemistry of combustion, and is recommended reading for the
beginner. The earlier work of \citet{Emmons1963,Emmons1966} and \citet{Lee1972}
provides a sound background on the advances made during the post-war boom era.
\citet{Grishin1997} provides an extensive review of the work conducted in
Russia in the 1970s, 80s and 90s.

This particular paper will discuss those models based upon the fundamental
principles of the physics and chemistry of wildland fire behaviour. Later
papers in the series will discuss those models based upon observation of fire
behaviour and upon mathematical analogies to fire spread. As the laws of
physics are the same no matter the origin of the modeller, or the location of
the model, physical models are essentially based on the same rules and it is
only the implementation of those rules that differs in each model. A brief
discussion of the fundamentals of wildland fire behaviour covering the
chemistry and physics is given, followed by discussions of how these are
applied in physical models themselves. This is then followed by a discussion of
the quasi-physical models.

\section{Fundamentals of fire and combustion}

Wildland fire is the complicated combination of energy released (in the form of
heat) due to chemical reactions (broadly categorised as an oxidation reaction)
in the process of combustion and the transport of that energy to surrounding
unburnt fuel and the subsequent ignition of said fuel. The former is the domain
of chemistry (more specifically, \emph{chemical kinetics}) and occurs on the
scale of molecules, and the latter is the domain of physics (more specifically,
\emph{heat transfer} and \emph{fluid mechanics}) and occurs on scales ranging
from millimetres up to kilometres (Table \ref{Table:Scales}). It is the
interaction of these processes over the wide range of temporal and spatial
scales that makes the modelling of wildland fire behaviour a not inconsiderable
problem.

\citet[pg. 81]{Grishin1997} proposed five relative independent stages in the
development of a deterministic physical model of wildland fire behaviour:
\begin{enumerate}
  \item {Physical analysis of the phenomenon of wildland fire
  spread; isolation of the mechanism governing the transfer of
  energy from the fire front into the environment; definition of the
  medium type, and creation of a physical model of the phenomenon.}
  \item {Determination of the reaction and thermophysical properties
  of the medium, the transfer coefficients and structural parameters
  of the medium, and deduction of the basic system of equations with
  corresponding additional (boundary and initial) conditions.}
  \item {Selection of a method of numerical solution of the problem,
  and derivation of differential equations approximating the basic
  system of equations.}
  \item {Programming; test check of the program; evaluation of the
  accuracy of the difference scheme; numerical solution of the
  system of equations.}
  \item {Testing to see how well the derived results comply with the
  real system; their physical interpretation; development of new
  technical suggestions for ways of fighting wildland fire.}
\end{enumerate}

Clearly, stages one and two represent considerable hurdles and sources of
contention for the best method in which to represent the phenomenon of wildland
fire.  This section aims to provide a background understanding of the chemistry
and physics involved in wildland fire. However, it must be noted that even
though these fields have made great advances in the understanding of what is
going on in these processes, research is still very active and sometimes cause
for contention \citep{diBlasi1998}.

\subsection{Chemistry of combustion}

The chemistry of combustion involved in wildland fire is necessarily a complex
and complicated matter. This is in part due to the complicated nature of the
fuel itself but also in the range of conditions over which combustion can occur
which dictates the evolution of the combustion process.

\subsubsection{Fuel chemistry}

\begin{figure}[p]
\centering
  \includegraphics[width=10cm]{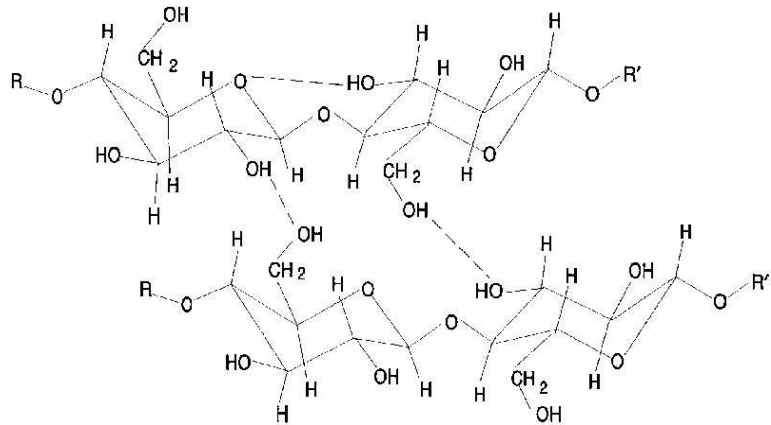}\\
  \caption{Schematic of chemical structure of portion of neighbouring cellulose
  chains, indicating some of the hydrogen bonds (dashed lines) that may stabilise the crystalline
  form of cellulose (Source: \citet{Ball1999}}\label{fig:cellulose}
\end{figure}

Wildland fuel is composed of live and dead plant material consisting primarily
of leaf litter, twigs, bark, wood, grasses, and shrubs. \citep{Beall1970}, with
a considerable range of physical structures, chemical components, age and level
of biological decomposition. The primary chemical constituent of biomass fuel
is cellulose (of chemical form  (C$_6$O$_5$H$_{10}$)$_n$), which is a polymer
of a glucosan (variant of glucose) monomer, C$_6$O$_6$H$_{12}$
\citep{Shafiz1982,Williams1982}. Cellulose is a linear, unbranched
polysaccharide of $\simeq$ 10,000 D-glucose units in $\beta (1,4)$
linkage\footnote{The D- prefix refers to one of two configurations around the
chiral centre of carbon-5. The $\beta (1,4)$ refers to the configuration of the
covalent link between adjacent glucose units, often called a glycosidic bond.
There are two possible geometries around C-1 of the pyranose (or 5-membered)
ring: in the $\beta$ anomer the hydrogen on C-1 sits on the opposite side of
the ring to that on C-2; in the $\alpha$ anomer it is on the same side. The
glycosidic bond in cellulose is between C-1 of one $\beta$ D-glucose residue
and the hydroxyl group on C-4 of the next unit (see Figure
\ref{fig:cellulose}).}. The parallel chains are held together by hydrogen
bonds, a non-covalent linage in which surplus electron density on hydroxyl
group oxygens is distributed to hydrogens with partial positive charge on
hydroxyl groups of adjacent residues \citep{Ball1999}.

Other major chemical components of wildland fuel include hemicelluloses
(copolymers of glucosan and a variety of other possibly monomers) and lignin (a
phenolic compound) in varying amounts, depending upon the species, cell type
and plant part (See Table \ref{Table:Fuel Percentages}). Minerals, water, salts
and other extractives and inorganics also exist in these fuels. The cellulose
is the same in all types of biomass, except for the degree of polymerisation
(i.e. the number of monomer units per polymer). Solid fuel is often referred to
as a \emph{condensed phase} fuel in the combustion literature.

Cellulose is an extraordinarily stable polysaccharide due to its structure:
insoluble, relatively resistant to acid and base hydrolysis, and inaccessible
to all hydrolytic enzymes except those from a few biological sources. Cellulose
is the most widely studied substance in the field of wood and biomass
combustion; by comparison, few studies have been carried out on the combustion
of hemicelluloses or lignin \citep{diBlasi1998}, due perhaps to the relative
thermal instability of these compounds. The degradation of biomass is generally
considered as the sum of the contribution of its main components (cellulose,
hemicelluloses and lignin) but the extrapolation of the thermal behaviour of
the main biomass components to describe the kinetics of complex fuels is only a
rough approximation \citep{diBlasi1998}. The presence of inorganic matter in
the biomass structure can act as a catalyst or an inhibitor for the degradation
of cellulose; differences in the purity and physical properties of cellulose
and hemicelluloses and lignin also play an important role in the degradation
process \citep{diBlasi1998}.

\subsubsection{Combustion reactions}

Chemical reactions can be characterised by the amount of energy required to
initiate a reaction, called the activation energy, $E_a$. This energy controls
the rate of reaction through an exponential relation, which can be derived from
first principles, known as the Arrhenius law:

\begin{equation}
k=A^{(\frac{-E_{a}}{RT})}\end{equation}

where \emph{k} is the reaction rate constant, \emph{A} is a pre-exponential
factor related to collision rate in Eyring theory, \emph{R} is the
gas constant and \emph{T} is the absolute temperature of the reactants.
Thus, the rate constant, \emph{k}, is a function of the temperature
of the reactants; generally the higher the temperature, the faster
the reaction will occur.

\subsubsection{Solid phase reactions--competing processes}

\begin{figure}[ht!]
  \centering
  \includegraphics[width=8cm]{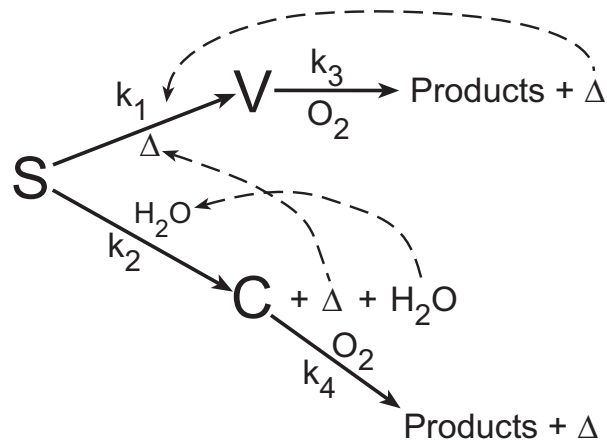}\\
  \caption{\small Schematic of the competing paths possible in the thermal degradation
  of cellulose substrate (S).  Volatilisation into levoglucosan (V) in the absence
  of moisture is endothermic.  Subsequent oxidisation of levoglucosan into products is
  exothermic.  Char formation (C) occurs at a lower activation energy in the presence of moisture.
  This path is exothermic and forms water.  Chemical and thermal feedback paths (dashed
  lines) can encourage either volatilisation or charring. (After \citet{diBlasi1998, Ball1999})}
  \label{Chemistry_paths}
\end{figure}

When heat is applied to cellulose, the cellulose undergoes a reaction called
thermal degradation. In the absence of oxygen, this degradation is called
\emph{pyrolysis}, even though in the literature the term pyrolysis is often
used incorrectly to describe any form of thermal degradation
\citep{Babrauskas2003}. Cellulose can undergo two forms of competing
degradation reaction: volatilisation and char formation (Figure
\ref{Chemistry_paths}). While each of these reactions involves the
depolymerisation of the cellulose (described as the `unzipping' of the polymer
into shorter strands \citep{Williams1982,Williams1985}), each has a different
activation energy and promoting conditions, and result in different products
and heat release.

\emph{Volatilisation} generally occurs in conditions of low or nil moisture and
involves thermolysis of glycosidic linkages, cyclisation and the release of
free levoglucosan via thermolysis at the next linkage in the chain
\citep{Ball2004}.  This reaction is endothermic (requiring about 300 J g$^{-1}$
\citep{Ball1999}) and has a relatively high activation energy (about 240 kJ
mol$^{-1}$\citep{diBlasi1998}). The product, levoglucosan (sometimes described
as `tar' \citep{Williams1982}), is highly unstable and forms the basis of a
wide range of subsequent species following further thermal degradation that
readily oxidise in the process of combustion, resulting in a multitude of
intermediate and final, gas and solid phase, products and heat.

\emph{Char formation}, on the other hand, occurs when thermal degradation
happens in the presence of moisture or low rates of heating. In this competing
reaction pathway, the nucleophile that bonds to the thermolysed carbo-cation at
C-1 is a water molecule. The initial  product is a reducing end which has `lost
the opportunity' to volatilise. Instead, further heating of such fragments
dehydrates, polyunsaturates, decarboxylates, and cross-links the carbon
skeleton of the structure, ultimately producing char. This reaction has a
relatively low activation energy (about 150 kJ mol$^{-1}$\citep{diBlasi1998})
and is exothermic (releasing about 1 kJ g$^{-1}$).

Thus, the thermal degradation of cellulose results in two competing pathways
controlled by thermal and chemical feedbacks such that if heating rates are low
and/or moisture is present, the charring pathway is promoted. If sufficient
energy is released in this process (or additional heat is added) or moisture
evaporated, then cyclisation and the release of free levoglucosan from
thermolysed, positively charged chain fragments, becomes statistically favoured
over nucleophilic addition of water and char production. If the subsequent
combustion of the levoglucosan and products releases enough energy then this
process becomes self-supporting. However, if the heat released is convected
away from the reactants or moisture is trapped, then the char-formation path
becomes statistically favoured. These two competing pathways will oscillate
until conditions become totally self-supporting or thermal degradation stops.

\subsubsection{Gas phase reactions}

Gas phase combustion of levoglucosan and its derivative products is
highly complex and chaotic. The basic chemical reaction is assumed
to be:

\begin{equation}
\textrm{C}_{6}\textrm{O}_{5}\textrm{H}_{10}+ 6\textrm{O}_{2} \rightarrow
6\textrm{C}\textrm{O}_{2}+5\textrm{H}_{2}\textrm{O},
\end{equation}

however, this assumes that all intermediate reactions, consisting of
oxidisation reactions of derivative products mostly, are complete. But the
number of pathways that such reactions can take is quite large, and not all
paths will result in completion to water and carbon dioxide.

As an example of a gas-phase hydrocarbon reaction, \citet{Williams1982} gives a
non-exhaustive list of 14 possible pathways for the combustion of CH$_4$, one
of the possible intermediates of the thermal degradation of levoglucosan, to
H$_2$O and CO$_{2}$. Intermediate species include CH$_3$, H$_2$CO, HCO, CO, OH
and H$_2$.

At any stage in the reaction process, any pathway may stop (through loss of
energy or reactants) and its products be advected away to take no further part
in combustion. It is these partially combusted components that form smoke.  The
faster and more turbulent the reaction, the more likely that reaction
components will be removed prior to complete combustion, hence the darker and
thicker the smoke from a headfire, as opposed to the lighter, thinner smoke
from a backing fire \citep{Cheney1997b}.

Because the main source of heat into the combustion process comes
from the exothermic reaction of the gas-phase products of
levoglucosan and these products are buoyant and generally convected
away from the solid fuel, the transport of the heat generated from
these reactions is extremely complex and brings us to the physics of
combustion.

\subsection{Physics of combustion}

The physics involved in the combustion of wildland fuel and the behaviour
of wildland fires is, like the chemistry, complicated and highly dependent
on the conditions in which a fire is burning. The primary physical
process in a wildland fire is that of heat transfer. \citet{Williams1982}
gives nine possible mechanisms for the transfer of heat from a fire:

\begin{enumerate}
\item Diffusion of radicals
\item Heat conduction through a gas
\item Heat conduction through condensed materials
\item Convection through a gas
\item Liquid convection
\item Fuel deformation
\item Radiation from flames
\item Radiation from burning fuel surfaces
\item Firebrand transport.
\end{enumerate}
1, 2 and 3 could be classed as diffusion at the molecular level. 4 and 5 are
convection (although the presence of liquid phase fuel is extremely rare) but
can be generalised to advection to include any transfer of heat through the
motion of gases. 7 and 8 are radiation. 6 and 9 could be classed as solid fuel
transport. This roughly translates to the three generally accepted forms of
heat transfer (conduction, convection and radiation) plus solid fuel transport,
which, as \citet{Emmons1966} points out, is not trivial or unimportant in
wildland fires.

The primary physical processes driving the transfer of heat in a wildland fire
are that of advection and radiation. In low wind conditions, the dominating
process is that of radiation \citep{Weber1989}. In conditions where wind is not
insignificant, it is advection that dominates \citep{Grishin1984b}. However, it
is not reasonable to assume one works without the other and thus both
mechanisms must be considered.

In attempting to represent the role of advection in wildland fire spread, the
application of fluid dynamics is of prime importance. This assumes that the gas
flow can be considered as a continuous medium or fluid.

\subsubsection{Advection or Fluid transport}

Fluid dynamics is a large area of active research and the basic outlines of the
principles are given here.  The interested reader is directed to a considerable
number of texts on the subject for more in-depth discussion (e.g.
\citet{Batchelor1967, Turner1973}).

The key aspect of fluid dynamics and its application to understanding the
motion of gases is the notion of continuity. Here, the molecules or particles
of a gas are considered to be \emph{continuous} and thus behave as a fluid
rather than a collection of particles.  Another key aspect of fluid mechanics
(and physics in general) is the fundamental notion of the conservation of
quantities which is encompassed in the fluidised \emph{equations of motion}.

A description of the rate of change of the density of particles in relation to
the velocity of the particles and distribution of particles provides a method
of describing the continuity of the particles. By taking the zeroth velocity
moment of the density distribution (multiplying by $\textbf{u}^k$ (where
\emph{k} = 0, in this case) and integrating with respect to \textbf{u}), the
equation of continuity is obtained. If the particles are considered to have
mass, then the continuity equation also describes the conservation of mass:

\begin{equation}
  \frac{\partial\rho}{\partial t}+\nabla.(\rho\mathbf{u})=0,\label{eq:Continuity}
\end{equation}

 where $\rho$ is density, \emph{t} is time, and \textbf{u} is the
fluid velocity (with vector components \emph{u}, \emph{v}, and \emph{w})
and $\nabla.$ is the Laplacian or gradient operator (i.e. in three
dimensions $\mathbf{i}\frac{\partial}{\partial x}+\mathbf{j}\frac{\partial}{\partial y}+\mathbf{k}\frac{\partial}{\partial z}$).
This is called the \emph{fluidised} form of the continuity equation
and is presented in the form of Euler's equations as a partial differential
equation.

However, in order to solve this equation, the evolution of \textbf{u} is
needed. This incompleteness is known as the closure problem and is a
characteristic of all the fluid equations of motion. The next order velocity
moment (\emph{k} = 1) can be taken and the evolution of the velocity field
determined. This results in an equation for the force balance of the fluid or
the \emph{conservation of momentum} equation:

\begin{equation}
  \frac{\partial\rho\mathbf{u}}{\partial t}+\nabla.(\rho\mathbf{u})\mathbf{u}+\nabla p = 0,\label{eq:Motion}
\end{equation}

where \emph{p} is pressure. However, the evolution of \emph{p} is
then needed to solve this equation.  This can be determined by
taking the second velocity moment (k=2) which provides an equation
for the conservation of energy, but it itself needs a further,
incomplete, equation to provide a solution.  One can either continue
determining higher order moments \emph{ad nauseum} in order to
provide a suitably approximate solution (as the series of equations
can never be truly closed) or, as is more frequently done, utilise
an equation of state to provide the closure mechanism. In fluid
dynamics, the equation of state is generally that of the ideal gas
law (e.g. $pV=nRT$). The above equations are in the form of the
Euler equations and represent a simplified (inviscid) form of the
Navier-Stokes equations.

\subsubsection{Buoyancy, convection and turbulence}

The action of heat release from the chemical reaction within the
combustion zone results in heated gases, both in the form of
combustion products as well ambient air heated by, or entrained
into, the combustion products.  The reduction in density caused by
the heating of the gas increases the buoyancy of the gas and results
in the gas rising as convection which can then lead to turbulence in
the flow. Turbulence acts over the entire range of scales in the
atmosphere, from the fine scale of flame to the atmospheric boundary
layer, and acts to mix heated gases with ambient air and to mix the
heated gases with unburnt solid phase fuels.  It also acts to
increase flame immersion of fuel. The action of turbulence also
affects the transport of solid phase combustion, such as that of
firebrands, resulting in spotfires downwind of the main burning
front.

Suitably formulated Navier-Stokes equations can be used to
incorporate the effects of buoyancy, convection and turbulence.
However, these components of the flow can be investigated
individually utilising particular approximations, such as
Boussinesq's concept of eddy viscosity for the modelling of
turbulence, or buoyancy as a renormalised variable for modelling the
effects of buoyancy. Specific methods for numerically solving
turbulence within the realm of fluid dynamics, including
renormalisation group theory (RNG) and large eddy simulation (LES),
have been developed. Convective flows are generally solved within
the broader context of the advection flow with a prescribed heat
source.

\subsubsection{Radiant heat transfer}

Radiant heat is a form of electromagnetic radiation emitted from a hot source
and is in the infra-red wavelength band. In flame, the primary source of the
radiation is thermal emission from carbon particles, generally in the form of
soot \citep{Gaydon1960}, although band emission from electronic transitions in
molecules also contributes to the overall radiation from a fire.

The general method of modelling radiant heat transfer is through the use of a
radiant transfer equation (RTE) of which the simplest is that of the
Stefan-Boltzmann equation:

\begin{equation}
  q=\sigma T^{4},
\end{equation}

where $\sigma$ is the Stefan-Boltzmann constant
($5.67\times10^{-8}$J\,K$^{-4}$\, m$^{-2}$\,s$^{-1}$ and \emph{T} is the
radiating temperature of the surface (K). While it is possible to approximate
the radiant heat flux from a fire as a surface emission from the flame face,
this does not fully capture the volumetric emission nature of the flame
\citep{Sullivan2003a} and can lead to inaccuracies in flux estimations if
precise flame geometry (i.e. view factor), temperature and emmissivity
equivalents are not known.

More complex solutions of the RTE, such as treating the flame as a volume of
radiation emitting, scattering and absorbing media, can improve the prediction
of radiant heat but are necessarily more computationally intensive; varying
levels of approximation (both physical and numerical) are frequently employed
to improve the computational efficiency. The Discrete Transfer Radiation Model
(DTRM) solves the radiative transfer equation throughout a domain by a method
of ray tracing from surface elements on its boundaries and thus does not
require information about the radiating volume itself. Discrete Ordinate Method
(DOM) divides the volume into discrete volumes for which the full RTE is solved
at each instance and the sum of radiation along all paths from an observer
calculated. The Differential Approximation (or P1 method) solves the RTE as a
diffusion equation which includes the effect of scattering but assumes the
media is optically thick. Knowledge about the media's absorption, scattering,
refractive index, path length and local temperature are required for many of
these solutions. Descriptions of methods for solving these forms of the RTE are
given in texts on radiant heat transfer (e.g. \citet{Drysdale1985}).
\citet{Sacadura2005} and \citet{Goldstein2006} review the use of radiative heat
transfer models in a wide range of applications.

Transmission of thermal radiation can be affected by smoke or band absorption
by certain components of the atmosphere (e.g CO$_2$, H$_2$O).

\subsubsection{Firebrands (solid fuel transport)}

Determination of the transport of solid fuel (i.e. firebrands),
which leads to the initiation of spotfires downwind of the main fire
front is highly probabilistic \citep{Ellis2000a} and not readily
amenable to a purely deterministic description. This is due in part
to the wide variation in firebrand sources and ignitions and the
particular flight paths any firebrand might take. Maximum distance
that a firebrand may be carried is determined by the intensity of
the fire and the updraught velocity of the convection, the height at
which the firebrand was sourced and the wind profile aloft
\citep{Albini1979, Ellis2000a}. Whether or not the firebrand lands
alight and starts a spotfire is dependent upon the nature of the
firebrand, how it was ignited, its combustion properties (including
flaming lifetime) and the ignition properties of the fuel in which
it lands (e.g. moisture content, bulk density, etc)
\citep{Plucinski2003}.

\subsubsection{Atmospheric interactions}

The transport of the gas phase of the combustion products interacts
with the atmosphere around it, transferring heat and energy, through
convection and turbulence. The condition of the atmosphere,
particularly the lapse rate, or the ease with which heated parcels
of air rise within the atmosphere, controls the impact that buoyancy
of the heated air from the combustion zone has on the atmosphere and
the fire.

Changes in the ambient meteorological conditions, such as changes in
wind speed and direction, moisture, temperature, lapse rates, etc,
both at the surface and higher in the atmosphere, can have a
significant impact on the state of the fuel (moisture content), the
behaviour of a fire, its growth, and, in turn, the impact that fire
can have on the atmosphere itself.

\subsubsection{Topographic interactions}

The topography in which a fire is burning also plays a part in the way in which
energy is transferred to unburnt fuel and the ambient atmosphere. It has long
been recognised that fires burn faster upslope than they do down, even with a
downslope wind. This is thought to be due to increased transfer of radiant heat
due to the change in the geometry between the fuel on the slope and the flame,
however recent work \citep{Wu2000} suggests that there is also increased
advection in these cases.

\section{Physical Models}

This section briefly describes each of the physical models that were
developed since 1990 (Table \ref{Phys_Sum}). Many are based on the
same basic principles and differ only in the methodology of
implementation or the purpose of use. They are presented in
chronological order of first publication. Some have continued
development, some have been implemented and tested against
observations, others have not. Many are implemented in only one or
two dimensions in order to improve computational or analytical
feasibility. Where information about the performance of the model on
available computing hardware is available, this is given.

\subsection{Weber (1991)}

Weber's \citeyearpar{Weber1991b} model was an attempt to provide the framework
necessary to build a physical model of fire spread through wildland fuel,
rather than an attempt to actually build one. To that end, Weber highlights
several possible approaches but does not give any definitive answer.

Weber begins with a reaction-transport formulation of the
conservation of energy equation, which states that the rate of
change of enthalpy per unit time is equal to the spatial variation
of the flux of energy plus heat generation. He then formulates
several components that contribute to the overall flux of energy,
including radiation from flames, radiation transfer to fuel through
the fuel, advection and diffusion of turbulent eddies. Heat is
generated through a chemical reaction that is modelled by an
Arrhenius law which includes heat of combustion.

This results in a first cut model that is one dimensional in \emph{x} plus
time. Advection, radiation and reaction components allow the evolution of the
fluid velocity to be followed. Solid phase and gas phase fuel are treated
separately due to different energy absorption characteristics.

In a more realistic version of this model, Weber treats the phase differences
more explicitly, producing two coupled equations for the conservation of
energy. The coupling comes from the fact that when the solid volatilises it
releases flammable gas that then combusts, returning a portion of the released
energy back to the solid for further volatilisation.

Weber determines that in two dimensions, the solution for the simple model is a
two-dimensional travelling wave that produces two parametric equations for
spatial \emph{x} and \emph{y} that yields an ellipse whose centre has been
shifted. Weber favourably compares this result with that of
\citet{Anderson1982}, who first formalised the spread of a wildland fire
perimeter as that of an expanding ellipse. No performance data are given.

\subsection{AIOLOS-F (CINAR S.A., Greece)}

AIOLOS-F was developed by CINAR S.A., Greece, as a decision support tool for
wildland fire behaviour prediction. It is a computational fluid dynamics model
that utilises the 3-dimensional form of the conservation laws to couple the
combustion of a fuel layer with the atmosphere to model forest fire spread
\citep{Croba1994}. It consists of two components, AIOLOS-T which predicts the
local wind field and wind-fire interaction, and AILOS-F which models the fuel
combustion.

The gas-phase conservation of mass equation is used to calculate the local wind
perturbation potential, the gas-phase conservation of momentum is used to
determine the vertical component of viscous flow, and a state equation to
predict the air density and pressure change with air temperature
\citep{Lymberopoulos1998}.

The combustion model is a 3D model of the evolution of enthalpy from which
change in solid-phase temperature is determined. A thermal radiation heat
transfer equation provides the radiant heat source term. Fuel combustion is
modelled through a 3-dimensional fuel mixture-fraction evolution that is tied
to a single Arrhenius Law for the consumption of solid phase fuel. The quantity
of fuel consumed by the fire within a time interval is an exponential function
of the mixture fraction.

The equations are solved iteratively and in precise order such that the wind
field is solved first, the enthalpy, mixture-fraction, and temperature second.
These are then used to determine the change in air density which is then fed
back into the wind field equations taking into account the change in buoyancy
due to the fire. The enthalpy, mixture-fraction and temperature are then
updated with the new wind field. This is repeated until a solution converges,
then the amount of fuel consumed for that time step is determined and the
process continues for the next time step.

Fuel is assumed to be a single layer beneath the lowest atmosphere grid. Fuel
is specified from satellite imagery on grids with a resolution in the order of
80 m. No data on calculation time is given, although it is described
\citep{Croba1994,Lymberopoulos1998} as being faster than real time.

\subsection{FIRETEC (Los Alamos National Laboratory, USA)}

FIRETEC \citep{Linn1997}, developed at the Los Alamos National
Laboratory, USA, is a coupled multiphase transport/wildland fire
model based on the principles of conservation of mass, momentum and
energy. It is fully 3-dimensional and in combination with a
hydrodynamics model called HIGRAD
\citep{Reisner1998,Reisner2000b,Reisner2000a}, which is used to
solve equations of high gradient flow such as the motions of the
local atmosphere, it employs a fully compressible gas transport
formulation to represent the coupled interactions of the combustion,
heat transfer and fluid mechanics involved in wildland fire
\citep{Linn2002b}.

FIRETEC is described by the author as self-determining, by which it
is meant that the model does not use prescribed or empirical relations
in order to predict the spread and behaviour of wildland fires, relying
solely on the formulations of the physics and chemistry to model the
fire behaviour. The model utilises the finite volume method and the
notion of a resolved volume to solve numerically its system of equations.
It attempts to represent the average behaviour of the gases and solid
fuels in the presence of a wildland fire. Many small-scale processes
such as convective heat transfer between solids and gases are represented
without each process actually being resolved in detail \citep{Linn1997,Linn1998a,Linn2002a}.
Fine scale wind patterns around structures smaller than the resolved
scale of the model, including individual flames, are not
represented explicitly.

The complex combustion reactions of a wildland fire are represented in FIRETEC
using a few simplified models, including models for pyrolysis, char burning,
hydrocarbon combustion and soot combustion in the presence of oxygen
\citep{Linn1997}. Three idealised limiting cases were used as a basis for the
original FIRETEC formulation:

\begin{enumerate}
\item gas-gas, with two reactants forming a single final product and no
intermediate species.
\item gas-solid, being the burning of char in oxygen.
\item single reactant, being pyrolysis of wood.
\end{enumerate}

However, \citet{Linn2002a} further refined this to a much simplified
chemistry model that reduced the combustion to a single solid-gas
phase reaction:

\begin{equation}
  N{}_{f}+N_{O_{2}}\rightarrow\, products+heat\label{eq:solid_reaction}
\end{equation}

where $N_{f,O_2}$ are the stoichiometric coefficients for fuel and oxygen. The
equations for the evolution of the solid phase express the conservation of
fuel, moisture and energy:

\begin{equation}
  \frac{{\partial\rho_{f}}}{\partial t}=-N_{f}F_{f}
\end{equation}

\begin{equation}
  \frac{{\partial\rho_{w}}}{\partial t}=-F_{w}
\end{equation}

\begin{eqnarray}
  (c_{p_{f}}\rho_{f}+c_{p_{w}}\rho_{w})\frac{\partial T_{s}}{\partial t} & = & Q_{rad,s}+ha_{v}(T_{g}-T_{s})-F_{w}(H_{w}+c_{p_{w}}T_{vap})+\nonumber \\
   &  & F(\Theta H_{f}+c_{p_{f}}T_{pyr}N_{f})
 \end{eqnarray}

where $F_{f,W}$ are the reaction rates for solid fuel and liquid
water depletion (i.e. the evaporation rate), $\rho_{f,w}$ are the
solid phase (i.e. fuel and liquid water) density, $\Theta$ is the
fraction of heat released from the solid-gas reaction and deposited
back to the solid, $c_{p_f,w}$ are the specific heats at constant
pressure of the fuel and water, $T_{s,g}$ is the temperature of the
solid or gas phase, $T_{pyr}$ is the temperature at which the solid
fuel begins to pyrolyse, $Q_{rad,g}$ is the net thermal radiation
flux to the gas, \emph{h} is the convective heat transfer
coefficient, $a_v$ is the ratio of solid fuel surface area to
resolved volume, $H_{w, f}$ is the heat energy per unit mass
associated with liquid water evaporation or solid-gas reaction (Eq.
\ref{eq:solid_reaction}). It is assumed that the rates of exothermic
reaction in areas of active burning are limited by the rate at which
reactants can be brought together in their correct proportions (i.e.
mixing limited). In a later work \citep{Colman2003} a procedure to
improve the combustion chemistry used in FIRETEC by utilising a
non-local chemistry model in which the formation of char and tar are
competing processes (as in for example, Fig. \ref{Chemistry_paths})
is outlined. No results have been presented yet.

The gas phase equations utilise the forms of the conservation of mass,
momentum, energy and species equations \citep{Linn2005b}, similar to those of
eqs (\ref{eq:Continuity} \& \ref{eq:Motion}), except that the conservation of
mass is tied to the creation and consumption of solid and gas phase fuel, a
turbulent Reynolds stress tensor and coefficient of drag for the solid fuel is
included in the momentum equation, and a turbulent diffusion coefficient is
included in the energy equation.

A unique aspect of the FIRETEC model is that the variables that occur in the
relevant solid and gas phase conservation equations are divided into mean and
fluctuating components and ensemble averages of the equations taken. This
approach is similar to that used for the modelling of turbulence in flows.

The concept of a critical temperature within the resolved volume is used to
initiate combustion and a probability distribution function based on the mean
and fluctuating components of quantities in the resolved volume used to
determine the mean temperature of the volume. Once the mean temperature exceeds
the critical temperature, combustion commences and the evolution equations are
used to track the solid and gas phase species. The critical temperature is
chosen to be 500 K \citep{Linn1997}.

Turbulence in the flow around the combusting fuel is taken into account as the
sum of three separate turbulence spectra corresponding to three cascading
spatial scales, \emph{viz.}: scale A, the scale of the largest fuel structure
(i.e. a tree); scale B, the scale of the distance between fuel elements (i.e.
branches); and scale C, the scale of the smallest fuel element (i.e. leaves,
needles, etc) \citep{Linn1997}. In the original work modelling fire spread
through a forest type, the characteristic scale lengths, \emph{s}, for each
scale were $s_{A}$ = 4.0 m, $s_{B}$ = 2.0 m and $s_{C}$ = 0.05 m. By
representing turbulence explicitly like this, the effect of diffusivity in the
transfer of heat can be included.

The original version of FIRETEC did not explicitly include the
effects of radiation, from either flame or fuel bed, or the
absorption of radiation into unburnt fuel--primarily because flames
and flame effects were at an unresolved scale within the model. As a
result fires failed to propagate in zero wind situations or down
slopes. Later revised versions
\citep{Bossert2000b,Linn2001,Linn2002a} include some form of radiant
transfer, however, this has not been formally presented anywhere and
\citet{Linn2003} admits to the radiant heat transfer model being
`very crude'.

Because FIRETEC models the conservation of mass, momentum and energy
for both the gas and solid phases, it does have the potential, via
the probability density function of temperature within a resolved
volume, to track the probability fraction of mass in a debris-laden
plume above the critical temperature \citep{Linn1998b} and thus
provide a method of determining the occurrence of `spotting'
downwind of the main fire.

Running on a 128-node SGI computer with R10000 processors, a simplified FIRETEC
simulation is described as running at `one to two orders of magnitude slower
than realtime' for a reasonable domain size \citep{Hanson2000}.

\subsection{Forbes (1997)}

\citet{Forbes1997} developed a two dimensional model of fire spread utilising
radiative heat transfer, species consumption and flammable gas production to
explain why most fires don't become major problems and why, when they do, they
behave erratically. The basis for his model is observations of eucalypt forest
fires which appeared either to burn quiescently or as raging infernos.

The main conceit behind the model is a two-path combustion model in which the
solid fuel of eucalypt trees either thermally degrades directly and rapidly in
an endothermic reaction, creating flammable fuel that then combusts
exothermically, or produces flammable `eucalypt vapours' endothermically which
then combust exothermically.

Forbes developed a set of differential equations to describe this process and,
because the reaction rates are temperature dependent, a temperature evolution
for both the solid and gas phases, which are the sum of radiation, conduction
(only included in the solid phase), convective heat loss, and the endothermic
reaction losses in the production of the two competing flammable gases. Wind is
included in the reaction equations.

Forbes concludes from his analysis of the one-dimensional form of
the equations that a travelling wave solution is only sustainable if
one of the two reaction schemes is endothermic overall and, since
this won't be the case in a large, intense bushfire, that bushfires
are unlikely to propagate as simple travelling waves. He determines
a solution of a one-dimensional line fire but found that for most
parameter values, the fire does not sustain itself. He found that
the activation energies for each reaction, rate constants and heat
release coefficients govern the propagation of the fire. Low
activation energies and temperatures and high heat release rates are
most likely to lead to growth of large fires.

Forbes then develops a two-dimensional solution for his equations,
making the assumption that the height of the processes involved in
the vertical direction (i.e. the flames) is small when compared to
the area of the fire (i.e. by some orders of magnitude). This
solution produces an elliptical fire shape stretched in the
direction of the wind. He suggests improving the model by including
fuel moisture. No performance data are given.

\subsection{Grishin (Tomsk State University, Russia)}

The work of AM Grishin has long been recognised for its
comprehensive and innovative approach to the problem of developing
physical models of forest fire behaviour \citep{Weber1991b}.  While
most of this work was conducted and published in Russia in the late
1970s and early 1980s, Grishin published a major monograph in 1992
that collected the considerable research he had conducted in one
place, albeit in Russian. In 1997, this monograph was translated
into English \citep{Grishin1997} (edited by Frank Albini) and, for
the first time, all of Grishin's work was available for English
readers and is the main reason for the inclusion of his work in this
review.

Grishin's model, as described in a number of papers
\citep{Grishin1983,Grishin1984a,Grishin1984b,Grishin2002}, was based
on analysis of experimental data and developed using the concepts
and methods of reactive media mechanics.  In this formulation, the
wildland fuel (in this case, primarily forest canopy) and combustion
products represent a non-deformable porous-dispersed medium
\citep{Grishin1997}.  Turbulent heat and mass transfer in the
forest, as well as heat and mass exchange between the near-ground
layer of the atmosphere and the forest canopy, are incorporated.
The forest is considered as a multi-phase, multi-storied, spatially
heterogeneous medium outside the fire zone.  Inside the fire zone,
the forest is considered to be a porous-dispersed, seven-phase,
two-temperature, single-velocity, reactive medium.  The six phases
within the combustion zone are: dry organic matter, water in liquid
state, solid products of fuel pyrolysis (char), ash, gas (composed
of air, flying pyrolytic products and water vapour), and particles
in the dispersed phase.

The model takes into account the basic physicochemical processes
(heating, drying, pyrolysis of combustible forest material) and
utilises the conservation of mass, momentum and energy in both the
solid and gas phases. Other equations, in conjunction with initial
and boundary conditions, are used to determine the concentrations of
gas phase components, radiation flux, convective heat transfer, and
mass loss rates through Arrhenius rate laws using
experimentally-determined activation energy and reaction rates.
Grishin uses an effective reaction whose mass rate is close to that
of CO to describe the combustion of `flying' pyrolytic materials,
because he determined that CO is the most common pyrolytic product
\citep{Grishin1983}. Numerical analysis then enables the structure
of the fire front and its development from initiation to be
predicted. Versions of the full formulation of the multi-phase model
are given in each of the works of Grishin (e.g. \citet{Grishin1983,
Grishin1997, Grishin2002}).

While the model is formulated for three spatial dimensions plus
time, the system of equations is generally reduced to a simpler form
in which the vertical dimension is averaged over the height of the
forest and the fire is assumed to be infinite in the y-direction,
resulting in a one-dimensional plus time system of equations in
which x is the direction of spread. The original formulation was
intended only for the acceleration phase from ignition until steady
state spread is achieved \citep{Grishin1983}.  This was extended
using a moving frame of reference and a steady-state rate of spread
(ROS) to produce an analytical solution for the ROS which was found
to vary linearly with wind speed \citep{Grishin1984a}.

The speed of the fire front is taken to be the speed of the 700 K isotherm. The
domain used for numerical analysis is in the order of 100-200 m long.  Rate of
spread is found to be dependent on initial moisture content of the fuel. No
performance data are given.

\subsection{IUSTI (Institut Universitaire des Syst\'emes Thermiqes Industriels,
France)}

IUSTI \citep{Larini1998,Porterie1998b,Porterie1998a,Porterie2000} is
based on macroscopic conservation equations obtained from local
instantaneous forms \citep{Larini1998} using an averaging method
first introduced by \citet{Anderson1967}. It aims to extend the
modelling approach of \citet{Grishin1983} to thermal non-equilibrium
flows. IUSTI considers wildland fire to be a multi-phase reactive
and radiative flow through a heterogeneous combustible medium,
utilising coupling through exchange terms of mass, momentum and
energy between a single gas phase and any number of solid phases.
The physico-chemical processes of fuel drying and pyrolysis due to
thermal decomposition are modelled explicitly.  Whereas FIRETEC was
intended to be used to model wildland fire spread across large
spatial scales, IUSTI concentrates on resolving the chemical and
conservation equations at a much smaller spatial scale at the
expense of 3-dimensional solutions. Thus, in its current
formulation, IUSTI is 2-dimensional in the x and z directions.

Having derived the set of equations describing the general analysis
of the reactive, radiative and multi-phase medium
\citep{Larini1998,Porterie1998b}, the authors of IUSTI then reduced
the system of equations to that of a much simplified version (called
a zeroth order model) in which the effects of phenomena were
dissociated from those of transfers. This was done by undertaking a
series of simplifying assumptions. The first assumption was that
solid particles are fixed in space, implying that solid phase
momentum is nil; there is no surface regression and no char
contribution in the conservation equations; and that the only
combustion process is that of pyrolysis in the gaseous phase. Mass
loss rates are deduced from Arrhenius-type laws following on from
the values used by \citet{Grishin1983} and \citet{Grishin1997} and
thermogravitic analysis \citep{Porterie2000}. Mass rate equations
for the conversion of solid fuel (gaseous production and solid fuel
mass reduction) assume an independent reaction path between char
formation and pyrolysis such that the rate of particle mass
reduction relative to thermal decomposition of the solid phase and
gas production rate is the sum of the all solid fuel mass loss rates
due to water vaporisation, pyrolysis, char combustion (as a
consequence of pyrolysis), and ash formation (as a consequence of
char oxidation from the idealised reaction, C $+$ O$_2 \rightarrow$
CO$_2$). The model also includes a set of equations governed by the
transition from the solid phase to a gas phase called the `jump'
condition because IUSTI considers such relatively small volumes. The
pyrolysis products are assumed to be removed out of the solid
instantaneously upon release. Mass diffusion of any chemical species
is neglected and no chemical reactions occur in the solid phase. A
single one-step reaction model in which fuel reacts with oxidant to
produce product is implemented.

A later version of IUSTI \citep{Porterie2000} utilises the
density-weighted or Favre average form of the conservation equations
due to the density variations caused by the heat release. The
time-averaged, density-weighted (Favre) fluctuation of turbulent
flux is approximated from Boussinesq's eddy viscosity concept and
the turbulent kinetic energy, $\kappa$, and dissipation rate,
$\epsilon$, are obtained from the renormalisation group theory
(RNG).

The formation of soot is modelled as the soot volume fraction which forms
mostly as a result of the pyrolysis process and so is assumed to be a
percentage of the mass loss rate due to pyrolysis. The radiative transfer
equation is based upon the mole fraction of the combustion products and the
average soot volume fraction, treating the gas as gray.

Drag is included through the drag coefficient which is a function of the
Reynolds number of the solid phase. Solid phase particles are treated as
spheres. The conductive/convective heat transfer coefficient is expressed using
the Reynolds number for flow around cylinders.

The governing equations of conservation in both gas and solid phases
are discretised on a non-uniform grid using a finite-volume scheme.
The domain over which the equations are solved is in the order of
1-2 m long by 0.1 m with an average resolution of $\simeq$ 0.01 m.

A one-dimensional version of this model was constructed in an attempt to
simplify the model \citep{Morvan2001b}.  A numerical experiment replicating
fire spread through a tube containing pine needles (in order to replicate
one-dimensional spread experimentally) was conducted.  Results showed a linear
increase in ROS with increasing wind speed up to a value of 16 cm s$^{-1}$.
Beyond this value, ROS dropped off dramatically and pyrolysis flow rate
reduced. Analysis of the species composition mass fractions showed that below
16 cm s$^{-1}$, the combustion is oxygen limited and is akin to smouldering
combustion. Above 16 cm s$^{-1}$, the combustion became fuel-limited as the
increased air flow increased convective cooling and slowed pyrolysis and hence
ROS.

No performance data are given.

\subsection{PIF97}

The detailed work of \citet{Larini1998}, \citet{Porterie1998b} and
\citet{Porterie2000} provided the framework for the development of a
related model, named PIF97 by its authors
\citep{Dupuy1999,Morvan2001a}.  The aim of this work was to simplify
the multi-phase IUSTI model of \citet{Larini1998} and
\citet{Porterie1998b} in order to develop a more
operationally-feasible model of wildland fire spread. The full 2D
IUSTI was reduced to a quasi-two-dimensional version in which the
fuel bed is considered to be one-dimensional and the gas
interactions (including radiation and convective mixing above the
bed) are two dimensional (x and z). In a manner similar to IUSTU,
two phases of media are considered--gas and solid. However, PIF97
assumes that the solid is homogeneous, unlike IUSTI which considers
multiple solid phases.

PIF97 comes in two parts. The first is a combustion zone model that
considers the radiative and convective heat transferred to the fuel
bed in front of the flaming zone. The radiative component considers
radiation flux from adjacent fuels, the ignition interface, flame
and the ambient media surrounding the fuel.  Radiation from solids
is assumed to be blackbody at a temperature of 1123K. This value was
selected so that the model could predict the spread of a single
experimental fire in pinaster needles.  Convective heat exchange
depends on the Nusselt number which is approximated through a
relation with the Reynolds number for the type of flow the authors
envisage. This in turn relies on the assumption of flow around a
cylinder of infinite length. Mass transfer and drag forces are
similarly derived using approximations to published models and
empirical correlations (i.e. assuming cylindrical particles). An
ignition temperature for solid fuel of 600 K is used.

The second part of the model is the fire-induced flow in the flaming
combustion zone behind the ignition interface. This depends on the
ROS of the interface derived from the combustion part of the model.
The temperature of this gas is assumed to be fixed at 900K. Other
values between 750K and 1050K were investigated but no significant
difference in results was found.

The numerical solution of PIF97 is based on a domain that is 25 cm
long and uses a spatial resolution of 1 mm. Results of the model are
compared to experimental results presented by \citet{Dupuy2000} in
which two radiation-only models, that of \citet{deMestre1989} and a
one-dimensional version of \citet{Albini1985,Albini1986a}  were
compared to laboratory experiments conducted with \emph{Pinus
pinaster} and \emph{P. halipensis} needles. PIF97 was found to be
comparable to the Albini model, except in \emph{P.halipensis}
needles where it performed markedly better. However, no model was
found to `perform well' in conditions of wind and slope.

A later version of PIF97 \citep{Morvan2004b} was extended to multiple solid
phases in order to simulate Mediterranean fuel complexes comprising live and
dead components of shrub and grass species, including twigs and foliage.  An
empirical correlation is used for the drag coefficient based on regular shapes
(i.e. cylinder, sphere, etc.) A RNG $\kappa-\epsilon$ turbulence model using
turbulent diffusion coefficients is incorporated and a pressure correction
algorithm used to couple the pressure with the velocity.

The revised model was implemented as a 2D vertical slice through the fire front
as a compromise between the computational time and need to study the main
physical mechanisms of the fire propagation. 80 \ex\ 45 control volumes, each
10 cm \ex\ 3 cm were used, defining a domain 8 m by 1.35 m.  ROS was defined as
the movement of the 500K isotherm inside the pyrolysis front. ROS was compared
to other models and observations of shrub fires \citep{Fernandes2001} and did
not perform well. The authors summarise their model as producing a ROS
relationship for wind $<$ 3 m s$^{-1}$ as `an increasing function of wind
speed' and then say the ROS reaches a limiting value at a wind speed of about 5
m s$^{-1}$. The other models and observations showed either linear or power law
(exp $<$ 1.0) relationships.

\citet{Dupuy2005} added a crown layer to this model resulting in six
families of solid phase fuel: three for shrubs (leaves and two size
classes of twigs (0-2, 2-6 mm), one for grass, and two for the
overstorey \emph{P. halepensis} canopy (needles and twigs 2-6 mm).
This version implemented a combustion model based on Arrenhius-type
laws after \citet{Grishin1997}. Soot production (for the radiation
transfer) was assumed at 5\% of the rate of solid fuel pyrolysis.

The domain was 200 m \ex\ 50 m high with, at its finest scale, cells 0.25 m
\ex\ 0.025 m, average of 0.25 \ex\ 0.25 and largest 1.0 \ex\ 0.25 m. 200 s of
simulation took 48 hours on an Intel Pentium P4 2GHz machine.

\subsection{LEMTA (Laboratoire d'\'Energ\'etique et de M\'ecanique Th\'eorique et Appliqu\'ee, France)}

This comprehensive model, developed by \citet{Sero2002} of Laboratoire
d'\'{E}nergétique et de M\'ecanique Th\'eorique et Appliqu\'ee in France,
considers a two-phase model, gas and solid, in three regions of a forest--above
the forest, in the forest and below the ground--at three scales: microscopic
(plant cell solid/gas level), mesoscopic (branch and leaf level) and
macroscopic (forest canopy/atmosphere level). They identify but do not
investigate a fourth scale, that of the `gigascopic' or landscape level.

The combustion chemistry is simplified in that only gas-phase combustion is
allowed. Solid phase chemistry only considers pyrolysis to gas-phase volatile
fuel, char and tar. Soot production is not considered, nor is char combustion.
Gas phases include O$_2$, water, N$_2$, fuel and inert residue. Solid to gas
phase transitions are handled by interface jump relations.

Conservation of species mass, momentum and energy are derived for mesoscopic
gas and solid phase interactions. These are then averaged over the larger
macroscopic scale by using distribution theory and convoluting the equations to
macroscopic quantities. Extended irreversible thermodynamics is then used to
close the system of equations. Arguments about thermal equilibrium are used to
further reduce the non-equilibrium equations for temperature and pressure.

The system of equations are then further simplified using assumptions about the
nature of the fuel (at rest) and the size and interaction of the fuel particles
with the gas phase (i.e. no advection, pressure or porosity variations in the
solid phase). Drag is not included. Gas phase equations in the region above the
forest do not include solid phase particles and, since soot is not modeled,
cannot suitably describe radiant heat from flames.

\citet{Margerit2002} and \citet{Chetehouna2004} reduced \citet
{Sero2002} to two dimensions in order to produce a more
operationally-feasible fire spread model. \citet{Margerit2002}
achieved this through assumptions that the scale of the fuel to the
fire was such that the fuel could be considered a boundary layer and
the fire a one-dimensional interface between burning and unburnt
fuel on the surface. (i.e. the fuel is thin relative to the width of
the fire). A few assumptions are then made: there is no vertical
component in the wind, the solid and gas phases are in thermal
equilibrium, and the non-local external radiative heat flux is
blackbody. The resulting two-dimensional model is a
reaction-diffusion model similar in form to \citet{Weber1991b}.
Assumptions about the speed of chemical reactions are made to get
the pyrolysis occurring at an ignition temperature.

\citet{Chetehouna2004} further reduced the two-dimensional
reaction-diffusion equations of \citet{Margerit2002} by making some
simplifying assumptions about the evaporation and ignition of the
solid phase fuel. Here, fixed temperatures are used, 100 and
300\degr C respectively.  Five distinct heating stages are used,
each separated by the temperature of the fuel: 1) fuel heating to
100\degr C; 2) moisture evaporation at 100\degr C; 3) fuel heating
to ignition temperature; 4) combustion at 300\degr C; and 5) mass
loss due to chemical reactions and heat loss at flame extinction.

Separate equations with different boundary conditions are used for
each stage but only stages 1-3 are important for fire spread. The
equations for these stages are then non-dimensionalised and a
limiting parameter, the thermal conductivity in the solid phase, is
used as a parameter for variation. The equations are then solved as
an eigenvalue problem in order to determine the ROS for each stage.
Two flame radiation models are used to incorporate long distance
radiant heat flux from flames: \citet{deMestre1989} and the version
given by \citet {Margerit2002}. Rates of spread are similar for both
flame models and reduce with increasing thermal conductivity.
However, despite the fact that the authors say the models compare
well with experimental results, no results or comparisons are given.

The model is then simulated on a computer. It provides a circular
shape in no wind/no slope, and an elongated shape under wind. An
example burning in real terrain is shown but no discussion of its
performance against real fires is given. Mention is made of it
operating in real-time on a PC.

\subsection{UoS (University of Salamanca, Spain)}

\citet{Asensio2002} constructed a 2D model of fire spread that used radiation
as the primary mode of heat transfer but also incorporated advection of hot gas
and convective cooling of fuels. The model, described here as UoS, employed a
simplified combustion chemistry model (only two phases: gas and solid, and two
species: fuel and oxygen) and utilised only conservation of energy and species
mass. It is assumed combustion is fuel limited and thus only one species is
conserved. Arrhenius laws for fuel consumption are used. Turbulence is not
accounted for directly or explicitly, but a term for advection with a wind
velocity vector is included.

The model is of a form that explicitly includes convective heating,
radiation, chemical energy release and natural convection.
Non-dimensionalised forms of the system of equations are then
discretised into a finite element form for numerical computation.
The model is considered to be a first step and the authors aim to
link it to the Navier-Stokes equations for better incorporation of
turbulence.

\citet{Asensio2005} attempted to provide a link from the 2D surface fire spread
of UoS to a model of convection above the fire. The model starts with the
conservation of momentum equation and then makes hydrostatic assumptions about
the atmosphere. It then decomposes this 3D model into a 2D model with height
that is averaged over a layer of fixed thickness. An asymptotic model is then
formed and solved producing 2D streamfunctions and an equation for the velocity
on the surface (which can then be inserted directly into the original spread
model for the advection of heat around the fire).

No performance data are given.

\subsection{WFDS (National Institute of Safety Technology, USA)}

The Wildland Fire Dynamic Simulator (WFDS), developed by the US
National Institute of Safety Technology \citep{Mell2006}, is an
extension of the model developed to predict the spread of fire
within structures, Fire Dynamic Simulator (FDS). This model is fully
3D, is based upon a unique formulation of the equations of motion
for buoyant flow \citep{Rehm1978} and is intended for use in
predicting the behaviour of fires burning through
peri-urban/wildlands (what the authors call `Community-scale fire
spread' \citep{Rehm2003,Evans2003}). The main objective of this
model is to predict the progress of fire through predominantly
wildland fuel augmented by the presence of combustible structures.

WFDS utilises a varying computational grid to resolve volumes as low
as 1.6 m (x) $\times$ 1.6 m (y) $\times$ 1.4 m (z) within a
simulation domain in the order of 1.5 km$^2$ in area and 200 m high.
Outside regions of interest, the grid resolution is decreased to
improve computation efficiency.

\citet{Mell2006} give a detailed description of the WFDS formulated for the
specific initial case of grassland fuels, in which vegetation is not resolved
in the gas-phase (atmosphere) grid but in a separate solid fuel (surface) grid
(which the authors admit is not suitable for fuels in which there is
significant vertical flame spread and air flow through the fuel). In the case
presented, the model includes features such as momentum drag caused by the
presence of the grass fuel (modelled as cylinders) which changes over time as
the fuel is consumed. Mechanical turbulence, through the dynamic viscosity of
the flow through the fuel, is modelled as a subgrid parameter via a variant of
the Large Eddy Simulation (LES) method.

The WFDS assumes a two-stage endothermic thermal decomposition (water
evaporation and then solid fuel `pyrolysis').  It uses the temperature
dependent mass loss rate expression of \citet{Morvan2004b} to model the solid
fuel degradation and assumes that pyrolysis occurs at 127\degr C. Solid fuel is
represented as a series of layers which are consumed from the top down until
the solid mass reaches a predetermined char fraction at which point the fuel is
considered consumed.

WFDS assumes combustion occurs solely as the result of fuel gas and oxygen
mixing in stoichiometric proportion (and thus is independent of temperature).
Char oxidation is not accounted for.  The gas phase combustion is modelled
using the following stoichiometric relation:
\begin{equation}
\textrm{C}_{3.4}\textrm{H}_{6.2}\textrm{O}_{2.5}+3.7(\textrm{O}_2+3.76\textrm{N}_2)\rightarrow
3.4\textrm{CO}_2+3.1\textrm{H}_2\textrm{O}+13.91\textrm{N}_2
\end{equation}

Due to the relatively coarse scale of the resolved computation grids within
WFDS, detailed chemical kinetics are not modelled.  Instead, the concept of a
mixture fraction within a resolved volume is used to represent the mass ratio
of gas-phase fuel to oxygen using a fast chemistry or flame sheet model which
then provides the mass loss flux for each species. The energy release
associated with chemical reactions is not explicitly presented but is accounted
for by an enthalpy variable as a function of species. The model assumes that
the time scale of the chemical reactions is much shorter than that of mixing.

Thermal radiation transport assumes a gray gas absorber-emitter using the P1
radiation model for which the absorption coefficient is a function of the
mixture fraction and temperature for a given mixture of species.  A soot
production model is not used; instead it is an assumed fraction of the mass of
fuel gas consumed.

\citet{Mell2006} provides simulation information for two experimental
grassfires.  In the first case, a high intensity fire in a plot 200 m $\times$
200 m within a domain of 1.5 km $\times$ 1.5 km and vertical height of 200 m
for a total 16 million grid cells, the model, running on 11 processors, took 44
cpu hours for 100 s of simulated time.  Another lower intensity experiment over
a similiar domain took 25 cpu hours for 100 s of simulated time.

\section{Quasi-physical models}

This section briefly describes quasi-physical models that have
appeared in the literature since 1990 (\citet{deMestre1989} is
included because it was missed by previous reviews and provides the
basis for a subsequent model).

The main feature of this form of model is the lack of combustion
chemistry and reliance upon the transfer of a prescribed heat
release (i.e. flame geometry and temperature are generally assumed).
They are presented in chronological order of publication (Table
\ref{Quas_Sum}).

\subsection{Australian Defence Force Academy (ADFA) I, Australia}

\citet{deMestre1989} of the Australian Defence Force Academy, University of New
South Wales, developed a physical model of fire spread based initially only on
radiative effects, in much the same manner as that of \citet{Albini1985,
Albini1986a} (see below) but in a much simplified manner.

The authors utilise a conservation of heat approach to model the
spread of a planar fire of infinite width across the surface of a
semi-transparent fuel bed in a no wind, no slope situation. However,
unlike Albini, who modelled the fuel bed in two dimensions (i.e., x
and z), \citet{deMestre1989} chose to model only the top of the fuel
bed, arguing that it is this component of the fuel bed that burns
first before burning down into the bed; thus this model is one
dimensional plus time.

Assumptions include vertical flames that radiate as an opaque
surface of fixed temperature and emissivity, a fixed fuel ignition
temperature, and radiation from the combustion zone as an opaque
surface of fixed temperature and emissivity. Here they also assume
that the ignition interface in the fuel bed is a linear surface, as
opposed to Albini's curved one, in order to simplify the
computations.

Two approaches to the vaporisation of the fuel moisture are modelled--one in
which it all boils off at 373 K (i.e. 3 phase model ($<$373 K, 373 K, $>$373 K)
and one in which it boils off gradually below 373 K (2 phase model ($\leq$ 373
K, $>$ 373 K).

The final model includes terms for radiation from flame, radiation from
combustion zone, radiative cooling from solid fuel, and convective cooling from
solid fuel. Without the cooling terms, the model was found to over-predict ROS
by a factor of 13. A radiative cooling factor brought the over-prediction down
to a factor of 9. Including a convective cooling term to the ambient air
apparently brought the prediction down to the observed ROS but this was not
detailed.

No performance data are given.

\subsection{TRW, USA}

\citet{Carrier1991} introduced an analytical model of fire spread
through an array of sticks in a wind tunnel (called here TRW).
Unlike many preceding fire modelling attempts, they did not assume
that the dominant preheating mechanism is radiation, but a mixture
of convective/diffusive (what they called `confusive') heating.

Predominately concerned with deriving a formula for the forward
spread of the fire interface in the wind tunnel (based on a series
of experiments conducted and reported by \citet{Wolff1991}),
\citet{Carrier1991} assumed that the fire achieves a `quasi-steady'
ROS in conditions of constant wind speed and fuel conditions. They
make the point that, at the scale they are looking at, the spread
can be viewed as continuous and can thus involve a catch-all heat
transfer mechanism (gas-phase diffusion flame) in which radiation
plays no part and it is the advection of hot gas from the burned
area that preheats the fuel (assuming all of it is burnt).

The model is two-dimensional in the plane XY in which it is assumed there is no
lateral difference in the spread of the fire (which is different to assuming an
infinite width fire). Indeed, their formulation actually needs the width of the
fuel bed \emph{and} the width of the wind tunnel. The fluctuating scale of the
turbulence within the tunnel is incorporated in a scale length parameter. Air
flow within the fuel bed is ignored.

Using a first-principles competing argument, they say that if radiation was to
be the source of preheating, the estimate of radiant energy (2.9 J/g) ahead of
the fire falls well short of the 250 J/g required for pyrolysis.  A square root
relation between wind speed normalised by fuel load consumed and rate of
forward spread was determined and supported by experimental observation
\citep{Wolff1991}.  \citet{Carrier1991} suggest that only when fuel loading is
very high (on the order of 2 kg m$^{-2}$) will radiative preheating play a role
comparable to that of convective/diffusive preheating.

No performance data are given.

\subsection{Albini, USA}

\citet{Albini1985,Albini1986a} developed a two-dimensional (XZ)
quasi-physical radiative model of fire spread through a single
homogeneous fuel layer. The latter paper improved upon the former by
including a fuel convective cooling term. Both models required that
flame geometry and radiative properties (temperature and emissive
power) be prescribed \emph{a priori} in order for the model to then
determine, in an iterative process, the steady-state speed of the
fire front. The fire front is considered to be the isothermal flame
ignition interface between unburnt and burnt fuel expressed as an
eigenvalue problem, utilising a 3-stage fuel heating model ($<$373
K, 373 K, 373 K $\leq$ T$_i$), where T$_i$ is the ignition
temperature of 700 K.

A modified version of this spread model, in which a thermally-inert
zone that allowed the passage of a planar flame front but did not
influence the spread process was placed beneath the homogeneous fuel
layer to simulate propagation of a fire through the tree crowns, was
tested against a series of field-based experimental crown fires
conducted in immature Jack Pine \citep{Albini1986b}. The results
from one experimental fire were used to parameterise the model
(flame radiometric temperature and free flame radiation) and obtain
flame geometry and radiative properties for the remaining fires. The
model was found to perform reasonably well, with a maximum absolute
percent error of 14\%.

\citet{Albini1996} extended the representation of the fuel to
multiple levels, where surface fuel, sub-canopy fuel and the canopy
fuel are incorporated into the spread model. Albini also introduced
a closure mechanism for removing the requirement for flame geometry
and radiative properties to be given \emph{a priori}.  The former
transcribed the fuel complex into a vertical series of equivalent
single-component (homogeneous) surrogate layers based on weighted
contributions from different fuel components (e.g. surface-volume
ratio, packing ratio, etc.) within a layer.

The closure method involves the positing of a `trial' rate of
spread, along with free flame geometry and ignition interface shape,
that are then used to predict a temperature distribution within the
fuel.  This distribution is then subsequently used in a sub-model to
refine the fire intensity, rate of spread, flame geometry, etc. This
continues iteratively until a convergence of results is achieved. A
quasi-steady ROS is assumed and the temperature distribution is
assumed stationary in time.  A conservation of energy argument, that
the ROS will yield a fire intensity that results in a flame
structure that will cause that ROS, is then used to check the
validity of the final solution.

\citet{Butler2004} used the heat transfer model of \citet{Albini1996} in
conjunction with models for fuel consumption, wind velocity profile and flame
structure, to develop a numerical model for the prediction of spread rate and
fireline intensity of high intensity crown fires.  The model was found to
accurately predict the relative response of fire spread rate to fuel and
environment variables but significantly over-predicted the magnitude of the
speed, in one case by a factor of 3.5. No performance data are given.

\subsection{University of Corsica (UC), France}

The University of Corsica undertook a concerted effort to develop a physical
model of fire spread (called here UC) that would be suitable for faster than
real-time operational use.  The initial approach
\citep{Santoni1998a,Santoni1998b,Balbi1999} was a quasi-physical model in which
the main heat transfer mechanisms were combined into a so-called `reactive
diffusion' model, the parameters of which were determined experimentally.

The main components of UC are a thermal balance model that
incorporates the combined diffusion of heat from the three
mechanisms and a diffusion flame model for determination of radiant
heat from flames.  The heat balance considers: heat exchanged with
the air around a fuel cell, heat exchanged with the cell's
neighbours, and  heat released by the cell during combustion.  It is
assumed that the  rate of energy release is proportional to the fuel
consumed and that the fuel is consumed exponentially. The model is
two-dimensional in the fuel layer (the plane XY). No convection,
apart from convective cooling to neighbouring cells is taken into
account, nor is turbulence. The model assumes that all combustion
follows one path.  Model parameters were determined from laboratory
experiments.

Initially, radiation from the flame was assumed to occur as surface
emission from a flame of height, angle and length computed from the
model and an isothermal of 500K. Flame emissivity and fuel
absorptivity were determined from laboratory experiments in a
combined parameter. The early version of the model was one
dimensional for the fuel bed and two dimensional (x and z) for
flame. Forms of the conservation of mass and momentum equations are
used to control variables such as gas velocity, enthalpy, pressure
and mass fractions.

\citet{Santoni1999} presented a 2D version of the model in which the radiative
heat transfer component was reformulated such that the view factor, emissivity
and absorptivity were parameterised with a single value for each fuel and slope
combination that was derived from laboratory experiments.  This version was
compared to experimental observations \citep{Dupuy1995} and the radiation-only
models of \citet {Albini1985,Albini1986a} and \citet{deMestre1989}
\citep{Morandini2000}. It was found to predict the experimental increase in ROS
with increasing fuel load much better than the other models. The UC model also
outperformed the other models on slopes but this is not surprising as it had to
be parameterised for each particular slope case.

\citet{Simeoni2001a} acknowledged the inadequacies of the initial
`reaction-diffusion' model and
\citet{Simeoni2001a,Simeoni2001b,Simeoni2002a,Simeoni2003} undertook
to improve the advection component of the UC model by reducing the
physical modelling of the advection component of the work of
\citet{Larini1998} and \citet{Porterie1998b, Porterie1998a} to two
dimensions to link it to the UC model. It occurred in two parts: one
as a conservation of momentum component that is included in the
thermal balance equations (temperature evolution), and one as a
velocity profile through the flaming zone. They assumed that
buoyancy, drag and vertical variation are equivalent to a force
proportional to the quantity of gas in the multi-phase volume and
that all the forces are constant whatever the gas velocity. The net
effect is that the horizontal velocity decreases through the flame
to zero at the ignition interface and does not change with time.
Again, the quasi-physical model was parameterised using a
temperature-time curve from a laboratory experiment with no wind or
slope.  The modified model improves the performance only marginally,
particularly in the no slope case but still underpredicts ROS.

The original UC model included only the thermal balance with a diffusion term
encompassing convection, conduction and radiation. The inclusion of only short
distance radiation interaction failed to properly model the pre-heating of fuel
due to radiation prior to the arrival of the fire front.
\citet{Morandini2001a,Morandini2001b,Morandini2002} attempted to address this
issue by improving the radiant heat transfer mechanism of the model. Surface
emission from a vertical flame under no wind conditions is assumed and a flame
tilt factor is included when under the influence of wind. The radiation
transfer is based on the Stefan-Boltzmann radiation transfer equation where the
view factor from the flame is simplified to the sum of vertical panels of given
width. The length of each panel is assumed to be equal to the flame depth,
mainly because flame height is not modelled.

In cases of combined slope and wind, it was assumed that the effects on flame
angle are independent of slope. \citet{Morandini2002} approximate the effects
of wind speed by an increase in flame angle in a manner similar to terrain
slope by taking the inverse tangent of the flame angle of a series of
experiments divided into the mid-flame wind speed. This is then considered a
constant for a range of wind speeds and slopes.  Again, the model is
parameterised using a laboratory experiment in no-wind, no
slope conditions.

Results are given for a range of slopes (-15 to +15\degr) and wind speeds
(1,2,3 m s$^{-1}$). The prediction in no wind and slope is good, as are the
predictions for wind and no slope. The model breaks down when slope is added to
high wind (i.e. 3 m s$^{-1}$). Here, however, they determine that their model
only works for equivalent flame tilt angles (i.e. slope and wind angles) up to
40 degrees.

The model is computed on a fine-scale non-uniform grid using the
same methods as \citet{Larini1998}. In this case the smallest
resolution is 1 cm$^3$ and 0.1 seconds. On a Sun Ultra II, the model
took 114 s to compute 144 s of simulation.  When the domain is
reduced to just the fire itself, the time reduced to 18 s.

\subsection{ADFA II, Aust/USA}

\citet{Catchpole2002} introduced a much refined and developed version of ADFA
\citep{deMestre1989}, here called ADFA II. Like ADFA I, it is a heat balance
model of a fuel element located at the top of the fuel bed. The overall
structure of the model is the same, with radiative heating and cooling of the
fuel (from both the flames and the combustion zone), and convective heating and
cooling. It is assumed that the flame emits as a surface and they use
laboratory experiments to determine the emissive energy flux based on a
Gaussian vertical flame profile and a maximum flame radiant intensity.  It
assumes infinite width for the radiative emissions.

Combustion zone radiation is treated similarly. Byram's convective
number \citep{Byram1959a} and fireline intensity \citep{Byram1959b}
are used to determine flame characteristics (angle, height, length,
etc). Empirical models are used to determine gas temperature profile
above and within the fuel as well as maximum gas temperature, etc.
ADFA II utilises an iterative method to determine ROS, similar to
that of \citet{Albini1996}, assuming that the fire is at a
steady-state rate of spread.

No performance data are given.

\subsection{Coimbra (2004)}

The aim of \citet{Vaz2004} was not to develop a new model of fire spread as
such, but to combine the wide range of existing models in such a way as to
create a seamless modular quasi-physical model of fire spread that can be
tailor-made for particular situations by picking and choosing appropriate
sub-models. The `library' of models from which the authors pick and choose
their sub-models are classified as: heat sink models (including
\citet{Rothermel1972,Albini1985, Albini1986a, deMestre1989}), which consider
the conservation of energy aspects of fuel heating and moisture loss and
ignition criteria; heat flux models (including \citet{Albini1985, Albini1986a,
vanWagner1967}), which consider the net exchange of radiative energy between
fuel particles; and heat source models (including \citet{Thomas1967,
Thomas1971}), that consider the generation of energy within the combustion zone
and provide closure for the other two types of models.

The authors compared a fixed selection of sub-models against data gathered from
a laboratory experiment conducted on a fuel bed 2 m wide by 0.8 m long under
conditions of no wind and no slope.  The set of models was found to
underpredict ROS by 46\%. This was improved to 6\% when observed flame height
was used in the prediction. Predicted flame height, upon which several of the
sub-models depend directly, did not correspond with observations, regardless of
the combination of sub-models selected. Rather than producing a fire behaviour
prediction system that utilises the best aspects of its component models, the
result appears to compound the inadequacies of each of the sub-models.  None of
the three classes of sub-models consider advection of hot gases in the heat
transfer.

\section{Discussion and summary}

The most distinguishing feature of a fully physical model of fire
spread in comparison with one that is described as being
quasi-physical is the presence of some form of combustion chemistry.
These models determine the energy released from the fuel, and thus
the amount of energy to be subsequently transferred to surrounding
unburnt fuel and the atmosphere, etc., from a model of the
fundamental chemistry of the fuel and its combustion. Quasi-physical
models, on the other hand, rely upon a higher level model to
determine the magnitude of energy to be transferred and generally
require flame characteristics to be known \emph{a priori}.

Physical models themselves can be divided primarily into two streams; those
that are intended for operational or experimental use (or at least field
validation) and those that are purely academic exercises.  The latter are
characterised by the lack of follow-up work (e.g.
\citeauthor{Weber1991b}, \citeauthor{Forbes1997}, UoS), although it is possible
that components of such models may later find their way into models intended
for operational or experimental use. Sometimes the nature of the model itself
dictates its uses, rather than the authors' intention. Grishin, IUSTI, and
LEMTA were all formulated with the intention of being useful models of fire
spread but either due to the complex nature of the models, the reduced physical
dimensionality of the model or the restricted domain over which the model can
operate feasibly, the model has not and most probably will not be used
operationally.

The remaining physical models, AIOLOS-F, FIRETEC, PIF97 and WFDS have all had
extended and ongoing development and each are capable of modelling the
behaviour of a wildland fire of landscape scale (i.e. computational domains in
excess of $\simeq$ 100 m. However, in the effort to make this computationally
feasible, each model significantly reduces both the resolution of the
computational domain and the precision of the physical models implemented.

Each of these remaining physical models is also different from the
others in that efforts to conduct validation of their performance
against large scale wildland fires have been attempted. Difficulties
abound in this endeavour. As is the case with any field experiment,
it is very difficult to measure all required quantities to the
degree of precision and accuracy required by the models.  In the
case of wildland fires, this difficulty is increased by two or three
orders of magnitude.  Boundary conditions are rarely known and other
quantities are almost never measured at the site of the fire itself.
Mapping of the spread of wildland fires is haphazard and highly
subjective.

IUSTI and PIF97 undertook validation utilising laboratory
experiments of suitable spatial scales in which the number and type
of variables were strictly controlled.  In many laboratory
experiments, the standard condition is one of no wind and no slope.
While wildland fires in flat terrain do occur, it is very rare (if
not impossible) for these fires to occur in no wind.  The ability to
correctly model the behaviour of a fire in such conditions is only
one step in the testing of the model.  Both IUSTI and PIF97 (as well
as a number of the quasi-physical models discussed here) were found
wanting in conditions of wind and/or slope.

\citet{Morvan2004a} argues that purely theoretical modelling with no
regard for field observations is of less use than a field-based
model for one particular set of circumstances. Validation against
fire behaviour observed in artificial fuel beds under artificial
conditions is only half the test of the worth of a model.  The
importance of comparison against field observation is not to be
understated.  For regardless of the conditions under which a field
experiment (i.e. an experimental fire carried out in naturally
occurring, albeit modified, conditions) is conducted, it is the real
deal in terms of wildland fire behaviour and thus provides the
complete set of interactions between fire, fuel, atmosphere and
topography. Both \citet{Linn2005b} and \citet{Mell2006} identified
significant deficiencies within their models (FIRETEC and WFDS,
respectively) that only comparison against field observations could
have revealed.

Both FIRETEC and WFDS attempted validation against large scale experimental
grassland fires \citep{Cheney1993} and thus avoided many of the issues of
validation against wildfire observations. However, the issue of identifying the
source of discrepancy in such complex models is just as difficult as obtaining
suitable data against which to test the model.

Computationally feasible models can be either constructed from
simple models or reduced from complete models \citep{Sero2002} and
each of the preceding physical models are very much in the latter
category. Quasi-physical models are very much of the former but
suffer from the same difficulties in validation against large scale
fires. Of the quasi-physical models discussed here, only those of
Albini have been tested against wildland fires, the others against
laboratory experimental fires.

However, being constructed from simple models may make the quasi-physical
models less complex but does not necessarily make them any more computationally
feasible.  Table \ref{Summary} shows a summary of the scope, resolution and
computation time available in the literature for each of the models.  Not all
models have such information, concentrating primarily on the underlying basis
of the models rather than their computational feasibility.  But for those
models whose \emph{raison de tre} is to be used actively for fire management
purposes, computational feasibility is of prime concern.  Here models such as
AIOLOS-F, FIRETEC, WFDS and UC standout from the others because of their stated
aim to be a useful tool in fire management.

PIF97, WFDS and UC all give nominal computation times for a given
period of simulation.  Only UC, being a quasi-physical model reduced
from a more fundamental model is better than realtime.  PIF97 and
WFDS, using the current level of hardware, are all much greater than
realtime (in the order of 450 times realtime for WFDS on 11
processors \citep{Mell2006}.  FIRETEC is described as being `several
orders of magnitude slower than realtime'. FIRETEC, WFDS and UoS are
significantly different from the other physical models (and most of
the quasi-physical models for that matter) in that their resolutions
are significantly larger (in some cases by two orders of magnitude).
However, the time step used by FIRETEC (0.002 s) in the example
given, means that the gains to be made by averaging the computations
over a larger volume are lost in using a very short time interval.

The authors of FIRETEC are resolved to not being able to predict the
behaviour of landscape wildland fires and suggest that the primary
use of purely physical models of fire behaviour is the study of
fires under a variety of conditions in a range of fuels and
topographies in scenarios that are not amenable to field
experimentation.  This is a laudable aim, and in an increasingly
litigious social and political environment, may be the only way to
study large scale fire behaviour in the future, but this assumes
that the physical model is complete, correct, validated and
verified. \citet{Hanson2000} suggest that the operational fire
behaviour models of the future will be reduced versions of the
purely physical models being developed today.

It is obvious from the performance data volunteered in the
literature, that the current approaches to modelling fire behaviour
\emph{on the hardware available today} are not going to provide fire
managers with the tools to enable them to conduct fire suppression
planning based on the resultant predicted fire behaviour. The level
of detail of data (type and resolution of parameters and variables)
required for input into these models will not be generally available
for some time and will necessarily have a high degree of
imprecision.

The basis for fire behaviour models of operational use is unlikely
to be one of purely physical origin, simply because of the
computational requirements to solve the equations of motion at the
resolutions necessary to ensure model stability.  Approximations do
and will abound in order to improve computational feasibility and it
is these approximations that lessen the confidence users will have
in the final results.  Such approximations span the gamut of the
chemical and physical processes involved in the spread of fire
across the landscape; from the physical structure of the fuel
itself, the combustion chemistry of the fuel, the fractions of
species within a given volume, turbulence over the range of scales
being considered, to the chemical and thermal feedbacks within the
atmosphere.

It is most likely that for the foreseeable future operational models will
continue to be of empirical origin.  However, there may be a trend towards
hybrid models of a more physical nature as the physical and quasi-physical
models are further developed and refined.

\section{Acknowledgements}

I would like to acknowledge Ensis Bushfire Research and the CSIRO Centre for
Complex Systems Science for supporting this project; Jim Gould and Rowena Ball
for comments on the draft manuscript; and the members of Ensis Bushfire
Research who ably assisted in the refereeing process, namely Miguel Cruz,
Stuart Matthews and Grant Pearce.


\begin{table}[H!]
  \small
  \centering
  \caption{Outline of the major biological, physical and chemical components and processes occurring in a
   wildland fire and the temporal and spatial (vertical and horizontal) scales
   over which they operate.}\label{Table:Scales}
  \begin{tabular}{cccc}
 \\
    \hline
    Type & Time scale (s) & Vertical scale (m) & Horizontal scale (m) \\
    \hline
    Combustion reactions & 0.0001 - 0.01 & 0.0001 - 0.01 & 0.0001 - 0.01 \\
    Fuel particles & - & 0.001 - 0.01 & 0.001 - 0.01 \\
    Fuel complex & - & 1 - 20 & 1 - 100 \\
    Flames & 0.1 - 30 & 0.1 - 10 & 0.1 - 2 \\
    Radiation & 0.1 - 30 & 0.1 - 10 & 0.1-50 \\
    Conduction & 0.01 - 10 & 0.01 - 0.1 & 0.01 - 0.1 \\
    Convection & 1 - 100 & 0.1 - 100 & 0.1 - 10 \\
    Turbulence & 0.1 - 1000 & 1 - 1000 & 1 - 1000 \\
    Spotting & 1 - 100 & 1 - 3000 & 1 - 10000 \\
    Plume & 1 - 10000 & 1 - 10000 & 1 - 100 \\
    \hline
  \end{tabular}
\end{table}

\vspace{3cm}

\begin{table}[H]
\small \caption{Approximate analysis of some biomass species
\citep{Shafiz1982}.\label{Table:Fuel Percentages}} \centering
\begin{tabular}{ccccc}
\hline Species& Cellulose (\%)& Hemicellulose (\%)& Lignin (\%)& Other (\%)\\
\hline
Softwood    & 41.0 & 24.0 & 27.8 & 7.2  \\
Hardwood    & 39.0 & 35.0 & 19.5 & 6.5  \\
Wheat straw & 39.9 & 28.2 & 16.7 & 15.2 \\
Rice straw  & 30.2 & 24.5 & 11.9 & 33.4 \\
Bagasse     & 38.1 & 38.5 & 20.2 & 3.2  \\
\hline
\end{tabular}
\end{table}

\vspace{3cm}

\begin{table}
  \centering
  \small
  \caption{Summary of physical models (1990-present) discussed here.\ }\label{Phys_Sum}
  \begin{tabular}{cccccc}
    \hline
    Model    & Author     & Year     & Country   & Dimensions & Plane \\
    \hline
    Weber    & Weber                 & 1991 & Australia & 2 & XY \\
    AIOLOS-F & Croba \emph{et al.}   & 1994 & Greece    & 3 & - \\
    FIRETEC  & Linn                  & 1997 & USA       & 3 & - \\
    Forbes   & Forbes                & 1997 & Australia & 1 & X \\
    Grishin  & Grishin \emph{et al.} & 1997 & Russia    & 2 & XZ \\
    IUSTI    & Larini \emph{et al.}  & 1998 & France    & 2 & XZ \\
    PIF97    & Dupuy \emph{et al.}   & 1999 & France    & 2 & XZ \\
    LEMTA    & Sero-Guillaume \emph{et al.} & 2002 & France & 2(3) & XY \\
    UoS      & Asensio \emph{et al.} & 2002 & Spain    & 2 & XY \\
    WFDS     & Mell \emph{et al.}    & 2006 & USA       & 3 & - \\
    \hline
  \end{tabular}
\end{table}

\vspace{3cm}

\begin{table}[ht!]
  \centering
  \small
  \caption{Summary of quasi-physical models (1990-present) discussed here.\ }\label{Quas_Sum}
  \begin{tabular}{cccccc}
    \hline
    Model   & Author   & Year   & Country   & Dimensions & Plane \\
    \hline
    ADFA I  & de Mestre   & 1989 & Australia & 1 & X  \\
    TRW     & Carrier     & 1991 & USA       & 2 & XY \\
    Albini  & Albini      & 1996 & USA       & 2 & XZ \\
    UC      & Santoni     & 1998 & France    & 2 & XY \\
    ADFA II & Catchpole   & 1998 & Aust/USA  & 2 & XZ \\
    Coimbra & Vaz         & 2004 & Portugal  & 2 & XY \\
    \hline
  \end{tabular}
\end{table}

\pagebreak

\begin{landscape}
\begin{table}[ht!]
  \centering
  \small
  \caption{Summary of all models}\label{Summary}
  \begin{tabular}{cccccccccccc}
    \hline\\[-6pt]
    Model & No. & Domain Size & \multicolumn{4}{c}{Resolution (m)}                                    & CPU  & Simulation & Computation &
    Comment\\
    \cline{4-7}\\[-6pt]
          & Dimensions     & (x $\times$ y $\times$ z) & $\Delta$x & $\Delta$y & $\Delta$z & $\Delta$t & No. \& Type   &  Time(s) & Time (s)\\

    \hline
    \multicolumn{1}{l}{\emph{Physical}}\\
    Weber & 2 & ? & ? & - & - & ? & ?& ? & ? & \\
    AIOLOS-F & 3 & 10 $\times$ 10 $\times$ ? km & ? & ? & ? &  ? & ? & ? & ? & $<$ real time\\
    FIRETEC & 3 & 320 $\times$ 160 $\times$ 615 m & 2 m & 2 m & 1.5 m & 0.002 s & 128 nodes & ? & ? & $>>$ real time\\
    Forbes & 2 & ? & ? & ? & - & ? & ? & ? & ?\\
    Grishin & 2 & 50 $\times$ - $\times$ 12 m & ? & - & ? & ? & ? & ? & & 700 K isotherm\\
    IUSTI & 2 & 2.2 $\times$ - $\times$ 0.9 m & 0.02 & - & 0.09 & ?  & ? & ? & ? & 500 K isotherm\\
    PIF97 & 2 & 200 $\times$ - $\times$ 50 m  & 0.25 & & 0.25 & 1 s & P4 2GHz & 200 s & 48 h & 500 K isotherm\\
    LEMTA & 2 & ? & ? & ? & - & ? & `PC' & ? & ? & $\simeq$ real time\\
    UoS & 2 & ? & 1.875 m & 1.875 m & - & 0.25$\mu$s & ? & ? & ? \\
    WFDS & 3 & 1.5 $\times$ 1.5 $\times$ 0.2 km & 1.5 m & 1.5 m & 1.4 m & -& 11 nodes & 100 s & 25 h &  \\
    \multicolumn{1}{l}{\emph{Quasi-physical}}\\
    ADFA I & 1 & ? & ? & - & - & ? & ? & ? \\
    TRW & 2 & ? & ? & ? & - & ? & ? & ? & ? \\
    Albini & 2 & ? & ? & - & ? & ? & ? & ? & ? \\
    UC & 1 & 1 $\times$ 1 $\times$ - m & 0.01 m & 0.01 & 0.01 &0.01 s & Sun Ultra II & 144 s & 114 s & 500 K isotherm\\
    ADFA II & 2 & ? & ? & - & ? & ? & ? & ? & ? \\
    Coimbra & 2 & ? & ? & ? & ? & ? & ? & ? & ? \\
    \hline
  \end{tabular}
\end{table}
\end{landscape}

\end{document}